\documentclass{inprep}
\usepackage{amsmath,graphicx}
\usepackage[utf8x]{inputenc}
\usepackage[russian]{babel}

\begin{document}

\begin{center}
\Large\textsc{Институт ядерной физики им.~Будкера}\\[35mm]
А.Г.~Грозин\\[10mm]
Квантовый компьютер для чайников\\[20mm]
Препринт ИЯФ 2004-40\\
\vfill
\textsc{Новосибирск}\\[2mm]
2004
\end{center}

\thispagestyle{empty}
\newpage

\begin{center}
\large
\textbf{Квантовый компьютер для чайников}\\[3mm]
\textit{А.Г.~Грозин}\\[3mm]
Институт ядерной физики им.~Будкера\\
Новосибирск 630090\\[5mm]
\textsc{Аннотация}
\end{center}
\begin{quotation}
Введение в квантовые компьютеры, квантовую криптографию и квантовую телепортацию
для тех, кто никогда об этих вещах не слышал, но знает квантовую механику.
Ряд простых примеров подробно рассмотрен
с использованием эмулятора квантового компьютера QCL.
\end{quotation}
\vfill
\begin{flushright}
\copyright{} Институт ядерной физики им.~Будкера
\end{flushright}

\thispagestyle{empty}
\newpage

{
\parindent=0pt
\begin{tabular}{r}
\hline
\hspace{112mm}
\end{tabular}
\par
}

\vspace{50mm}

\section{Квантовый компьютер}

Я не буду обсуждать попытки технической реализации квантового компьютера.
Много групп во всём мире этим занимаются,
используя разнообразные подходы.
Достигнут некоторый прогресс, и, несомненно,
значительно большее продвижение произойдёт в близком будущем.
Но до работающего квантового компьютера, способного делать что-то полезное,
пока далеко.
Я не специалист по конкретным методам реализации.
Вместо этого, я расскажу о некоторых простых вещах,
которые можно будет делать с квантовым компьютером,
когда технические трудности его создания будут преодолены.

Литература на эту тему весьма обширна.
Я не пытался составить подробный список литературы;
приведены только ссылки на некоторые работы,
посвященные тем конкретным вопросам, которые будут обсуждаться.
Более обширные списки литературы можно найти в учебниках
и подробных обзорах.

\subsection{Устройство и система команд}

Главная часть квантового компьютера -- память, состоящая из квантовых бит.
Будем считать, что квантовый бит -- это спин $\frac{1}{2}$;
спин вверх означает $|0{>}$, вниз -- $|1{>}$.
Конечно, годится и любая другая система с двумя базисными состояниями.
Память из $n$ бит может быть в состоянии $|00000000{>}$
или $|01100001{>}$ или \dots{} или в суперпозиции таких базисных состояний.
Состояние памяти -- вектор в $2^n$-мерном пространстве.

\begin{figure}[ht]
\begin{center}
\includegraphics[width=\textwidth]{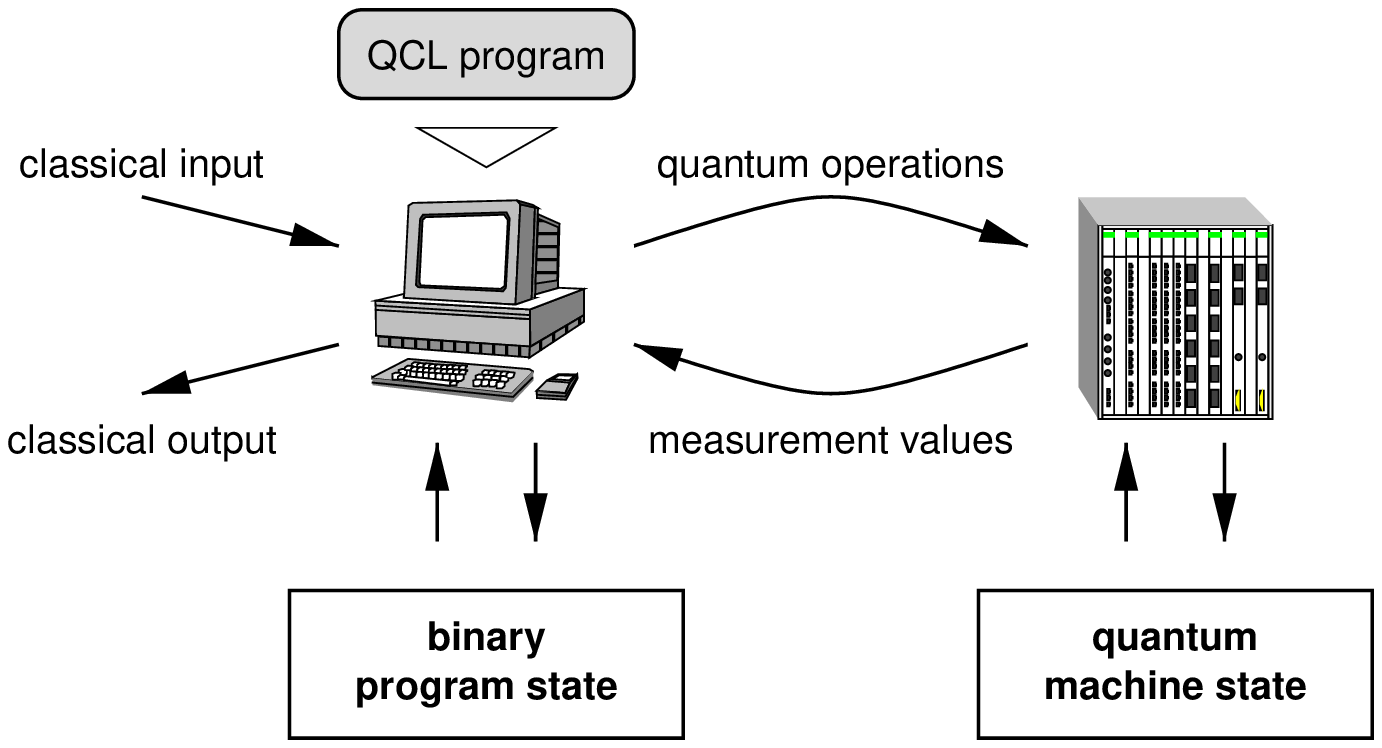}
\end{center}
\caption{Квантовый компьютер}
\label{Fig}
\end{figure}

Как известно, всякий эксперимент в квантовой механике
состоит из 3 этапов: приготовление начального состояния;
эволюция состояния (описываемая уравнением Шрёдингера);
измерение какой-нибудь наблюдаемой в конечном состоянии.
Цикл вычислений на квантовом компьютере
состоит из тех же 3 этапов:
\begin{itemize}
\item память приводится в начальное состояние $|00000000{>}$;
\item производится некоторая последовательность унитарных преобразований,
  действующих на отдельные биты или пары битов;
\item измеряется один или несколько битов.
\end{itemize}
Унитарные преобразования (команды квантового компьютера)
выполняются под управлением программы в обычном классическом компьютере.
Это показано на рис.~\ref{Fig}, который позаимствован из~\cite{Oemer}.

Законы квантовой механики ограничивают то,
какие операции мы можем производить с памятью
квантового компьютера во время вычислений.
Всякое унитарное преобразование обратимо
\[ \hat{U}^{- 1} = \hat{U}^+\,. \]
Поэтому во время работы квантового компьютера невозможны необратимые операции.
Нельзя, например, установить бит в 0, затерев его предыдущее значение.

Нельзя также скопировать бит в другой бит,
даже если второй бит находится перед этой операцией
в определённом состоянии, скажем, $|0{>}$.
Допустим, существует оператор $\hat{U}$
такой, что для всех $|\psi{>}$
\[ \hat{U} |\psi{>} \otimes |0{>} = |\psi{>} \otimes |\psi{>}\,. \]
Тогда
\[ \hat{U} |0{>} \otimes |0{>} = |0{>} \otimes |0{>}\,,\quad
\hat{U} |1{>} \otimes |0{>} = |1{>} \otimes |1{>}\,. \]
Для $|\psi{>} = \alpha |0{>} + \beta |1{>}$, по линейности,
\[ \hat{U} |\psi{>} \otimes |0{>} = \alpha |0{>} \otimes |0{>}
+ \beta |1{>} \otimes |1{>} \neq |\psi{>} \otimes |\psi{>}\,, \]
то есть такого оператора не существует.

Если некоторое вычисление $\hat{U}$ с начальным состоянием памяти $|i{>}$
даёт результат $\hat{U} |i{>}$,
то при начальном состоянии памяти $\sum_i \psi_i |i{>}$
оно даст результат $\sum_i \psi_i \hat{U} |i{>}$.
То есть все $2^n$ вычислений $\hat{U} |i{>}$ как бы производятся параллельно.
Это позволяет надеяться, что некоторые задачи можно решать
на квантовом компьютере гораздо быстрее, чем на классическом.
Но при измерении мы узнаём лишь один ответ,
причём мы не можем контролировать, какой именно.
Хорошо спроектированный квантовый алгоритм должен выдавать
желаемый ответ с вероятностью 1,
или хотя бы с вероятностью порядка 1
(тогда можно повторить вычисление несколько раз).
Если вероятность получения желаемого ответа экспоненциально мала,
то от такого алгоритма нет никакой пользы.

Лучше один раз увидеть, чем сто раз услышать.
Поскольку у нас нет работающего квантового компьютера,
мы будем использовать вместо него эмулятор qcl
(quantum computing language)~\cite{Oemer},
работающий на обычном классическом компьютере.
Это свободная программа, каждый может скачать её
и поэкспериментировать с ней.
Ниже приводится сеанс работы с qcl,
демонстрирующий команды квантового компьютера --
операции с квантовыми битами и их парами.%
\footnote{Этот сеанс работы производился из GNU \TeX{}macs~\cite{Hoeven} --
свободного wysiwyg текст-процессора, обеспечивающего прекрасное качество
не только текста, но и формул.
Я занимался его русификацией, а также написал интерфейсы \TeX{}macs
с рядом систем компьютерной алгебры~\cite{Grozin} и с qcl
(и при этом внёс ряд мелких улучшений в qcl).
Здесь использовалась версия qcl-0.6.1;
команда его вызова из \TeX{}macs была заменена на
\begin{flushleft}
\texttt{qcl -{}-texmacs -{}-dump-format=b -{}-auto-dump=32}
\end{flushleft}}

\noindent
\texttt{QCL Quantum Computation Language (32 qubits, seed 1081693763)}

\vspace{2mm}
\noindent
[0/32] $1\hspace{0.25em}|0{>}$

\vspace{2mm}
\noindent
Квантовый компьютер с 32 битами памяти запущен.
Назовём один бит \texttt{x}:

\noindent
\texttt{qcl> qureg x[1]}

\vspace{2mm}
\noindent
Поворот вокруг оси $x$:
\[ \hat{R}_x(\vartheta) = \left(\begin{array}{cc}
     \cos \frac{\vartheta}{2} & - i \sin \frac{\vartheta}{2}\\
     - i \sin \frac{\vartheta}{2} & \cos \frac{\vartheta}{2}
   \end{array}\right)\,. \]

\noindent
\texttt{qcl> RotX(pi/2,x)}

\noindent
[1/32] $0.70711\hspace{0.25em}|0{>} - 0.70711 i\hspace{0.25em}|1{>}$

\vspace{2mm}
\noindent
Поворот вокруг оси $y$:
\[ \hat{R}_y(\vartheta) = \left(\begin{array}{cc}
     \cos \frac{\vartheta}{2} & - \sin \frac{\vartheta}{2}\\
     \sin \frac{\vartheta}{2} & \cos \frac{\vartheta}{2}
   \end{array}\right)\,. \]

\noindent
\texttt{qcl> RotY(pi/2,x)}

\noindent
[1/32] $(0.5+0.5i)\hspace{0.25em}|0{>}
+ (0.5-0.5i)\hspace{0.25em}|1{>}$

\vspace{2mm}
\noindent
Поворот вокруг оси $z$:
\[ \hat{R}_z(\vartheta) = \left(\begin{array}{cc}
     e^{- i \vartheta / 2} & 0\\
     0 & e^{i \vartheta / 2}
   \end{array}\right)\,. \]

\noindent
\texttt{qcl> RotZ(pi,x)}

\noindent
[1/32] $(0.5-0.5i)\hspace{0.25em}|0{>}
+ (0.5+0.5i)\hspace{0.25em}|1{>}$

\vspace{2mm}
\noindent
Оператор Not:
\[ \hat{N} = i \hat{R}_x(\pi) = \left(\begin{array}{cc}
     0 & 1\\
     1 & 0
   \end{array}\right)\,,\quad
\hat{N} |0{>} = |1{>}\,,\quad
\hat{N} |1{>} = |0{>}\,. \]

\noindent
\texttt{qcl> Not(x)}

\noindent
[1/32] $(0.5+0.5i)\hspace{0.25em}|0{>}
+ (0.5-0.5i)\hspace{0.25em}|1{>}$

\vspace{2mm}
\noindent
Приведём память в исходное состояние:

\noindent
\texttt{qcl> reset}

\noindent
[1/32] $1\hspace{0.25em}|0{>}$

\vspace{2mm}
\noindent
Преобразование Адамара -- поворот на $\pi$ вокруг биссектрисы $x$ и $z$:
\[ \hat{M} = \frac{1}{\sqrt{2}}  \left(\begin{array}{cc}
     1 & 1\\
     1 & - 1
   \end{array}\right)\,. \]
Когда оно применяется к состоянию $|0{>}$,
получается
\[ |{+}{>} = \frac{|0{>} + |1{>}}{\sqrt{2}}\,: \]

\noindent
\texttt{qcl> Mix(x)}

\noindent
[1/32] $0.70711\hspace{0.25em}|0{>}
+ 0.70711\hspace{0.25em}|1{>}$

\vspace{2mm}
\noindent
Его квадрат равен 1:

\noindent
\texttt{qcl> Mix(x)}

\noindent
[1/32] $1\hspace{0.25em}|0{>}$

\vspace{2mm}
\noindent
Когда оно применяется к состоянию $|1{>}$,
получается
\[ |{-}{>} = \frac{|0{>} - |1{>}}{\sqrt{2}}\,: \]

\noindent
\texttt{qcl> Not(x)}

\noindent
[1/32] $1\hspace{0.25em}|1{>}$

\vspace{2mm}
\noindent
\texttt{qcl> Mix(x)}

\noindent
[1/32] $0.70711\hspace{0.25em}|0{>}
- 0.70711\hspace{0.25em}|1{>}$

\vspace{2mm}
\noindent
Опишем ещё 1 бит и назовём его \texttt{y}:

\noindent
\texttt{qcl> qureg y[1]}

\vspace{2mm}
\noindent
Повернём его вокруг оси $x$:

\noindent
\texttt{qcl> RotX(pi/2,y)}

\noindent
[2/32] $0.5\hspace{0.25em}|0,0{>}
- 0.5\hspace{0.25em}|1,0{>}
- 0.5i\hspace{0.25em}|0,1{>}
+ 0.5i\hspace{0.25em}|1,1{>}$

\vspace{2mm}
\noindent
Если каждый бит живет своей жизнью, это неинтересно:
тогда мы имеем $n$ независимых 1-битовых компьютеров
вместо одного $n$-битового.
Нужен оператор, действующий на пару битов.
Обычно в качестве него выбирают простейший -- Controlled Not.
Если первый (управляющий) бит равен 0,
то со вторым (управляемым) битом ничего не делается;
если управляющий бит равен 1,
то к управляемому биту применяется Not:
\begin{eqnarray*}
&& \hat{C} |00{>} = |00{>}\,,\quad \hat{C} |01{>} = |01{>}\,,\\
&& \hat{C} |10{>} = |11{>}\,,\quad \hat{C} |11{>} = |10{>}\,.
\end{eqnarray*}
Легко видеть, что это унитарная матрица $4\times4$.

\noindent
\texttt{qcl> x->y}

\noindent
[2/32] $0.5\hspace{0.25em}|0,0{>}
+ 0.5i\hspace{0.25em}|1,0{>}
- 0.5i\hspace{0.25em}|0,1{>}
- 0.5\hspace{0.25em}|1,1{>}$

\vspace{2mm}
\noindent
Теперь выключим квантовый компьютер:

\noindent
\texttt{qcl> exit}

Разумеется, когда у нас будет настоящий квантовый компьютер,
мы не сможем видеть состояние памяти после каждой операции.
Это возможно только для эмулятора.
Мы сможем что-то узнать только после измерения,
да и то лишь вероятностным образом.

\subsection{Пример квантового алгоритма}

В обычном классическом программировании функции одного параметра
реализуют так:
\begin{itemize}
\item параметр помещается во входной регистр;
\item команды, образующие тело функции, производят над ним
некие манипуляции, и помещают результат в выходной регистр,
затирая его прежнее состояние.
\end{itemize}
В квантовом программировании последняя операция невозможна,
так как она необратима.
Вместо этого, биты результата прибавляются
к битам выходного регистра по модулю 2.
Иными словами, над ними производится операция Xor
(исключающее или).
Эта операция, очевидно, обратима:
достаточно применить её второй раз,
и память вернётся в исходное состояние.

Рассмотрим функции, отображающие 1 бит в 1 бит:
\[ f : \{ 0, 1 \} \rightarrow \{ 0, 1 \}\,. \]
Каждой из них сопоставляется унитарный оператор
\[ \hat{U}_f |x, y{>} = |x, y \oplus f ( x ){>}\,. \]
Таких функций 4:
\begin{itemize}
  \item 2 константы: $f(x)=0$ и $f(x)=1$;
  \item 2 ``уравновешенные'' функции: $f(x)=x$ и $f(x)=1-x$
(они называются уравновешенными, потому что принимают значения 0 и 1
в равном числе точек).
\end{itemize}

\begin{sloppypar}
Допустим, кто-нибудь загадал функцию, а мы хотим угадать,
к которому из этих двух классов она принадлежит.
Для этого существует квантовый алгоритм~\cite{Deutsch},
который совершенно не похож на всё, к чему мы привыкли.
Как $\hat{U}_f$ действует на состояние $|0{>} \otimes |{-}{>}$,
где
\[ |{-}{>} = \frac{|0{>} - |1{>}}{\sqrt{2}}\,\mathrm{?} \]
Мы получим
\begin{eqnarray*}
&&\hat{U}_f |0,0{>} = |0,f(0){>}\,,\quad
\hat{U}_f |0,1{>} = |0,1-f(0){>}\,,\\
&&\hat{U}_f |0{>} \otimes |{-}{>} = \frac{|0,f(0){>} - |0,1-f(0){>}}{\sqrt{2}}
= (-1)^{f(0)} |0{>} \otimes |{-}{>}\,.
\end{eqnarray*}
Точно так же
\[ \hat{U}_f |1{>} \otimes |{-}{>} = (-1)^{f(1)} |1{>} \otimes |{-}{>}\,. \]
Теперь возьмём начальное состояние $|{+}{>} \otimes |{-}{>}$:
\begin{eqnarray*}
\hat{U}_f |{+}{>} \otimes |{-}{>} & = & \hat{U}_f
\frac{|0{>} \otimes |{-}{>} + |1{>} \otimes |{-}{>}}{\sqrt{2}}\\
& = & \frac{(-1)^{f(0)} |0{>} \otimes |{-}{>}
+ (-1)^{f(1)} |1{>} \otimes |{-}{>}}{\sqrt{2}}\,.
\end{eqnarray*}
Если $f(x)$ -- это константа, то $(-1 )^{f(1)} = (-1)^{f(0)}$,
и получается $(-1)^{f(0)} |{+}{>} \otimes |{-}{>}$.
Если $f(x)$ -- уравновешенная функция, то $(-1 )^{f(1)} = -(-1)^{f(0)}$,
и получается $(-1)^{f(0)} |{-}{>} \otimes |{-}{>}$.
Применим $\hat{M}$ к первому биту.
Для константы получится $|0{>}$,
а для уравновешенной функции $|1{>}$.
\end{sloppypar}

\vspace{2mm}
\noindent
\texttt{QCL Quantum Computation Language (32 qubits, seed 1081606808)}

\noindent
[0/32] $1\hspace{0.25em}|0{>}$

\vspace{2mm}
\noindent
Необходимые описания:

\noindent
\texttt{qcl> qureg x[1]; qureg y[1]; int r;}

\vspace{2mm}
\noindent
Это оператор $\hat{U}_f$:
\begin{itemize}
\item $n=0$ -- для $f(x)=0$ (этот оператор ничего не делает);
\item $n=1$ -- для $f(x)=1$;
\item $n=2$ -- для $f(x)=x$;
\item $n=3$ -- для $f(x)=1-x$.
\end{itemize}

\noindent
\begin{flushleft}
\texttt{qcl> procedure U(int n, qureg x, qureg y)}\\
\texttt{\ \ \ \ \ \{ if n==1 \{ Not(y); \}\ \ \ \ \ \ \ /* f(x)=1 */}\\
\texttt{\ \ \ \ \ \ \ else \{ if n==2 \{ x->y; \}\ \ /* f(x)=x */}\\
\texttt{\ \ \ \ \ \ \ else \{ if n==3 \{ Not(x); x->y; Not(x); \}\}\}\}}\\
\texttt{\ \ \ \ \ \ \ \ \ \ \ \ \ \ \ \ \ \ \ \ \ \ \ \ \ \ \ \ \ \ \ \ \ /* f(x)=1-x */}\\
\texttt{\ \ \ \ \ \}}
\end{flushleft}

\vspace{2mm}
\noindent
Приготовим состояние $|{-}{>}$ бита \texttt{y}:

\noindent
\texttt{qcl> Not(y)}

\noindent
[2/32] $1\hspace{0.25em}|0,1{>}$

\vspace{2mm}
\noindent
\texttt{qcl> Mix(y)}

\noindent
[2/32] $0.70711\hspace{0.25em}|0,0{>}
- 0.70711\hspace{0.25em}|0,1{>}$

\vspace{2mm}
\noindent
Теперь приготовим состояние $|{+}{>}\otimes|{-}{>}$:

\noindent
\texttt{qcl> Mix(x)}

\noindent
[2/32] $0.5\hspace{0.25em}|0,0{>}
+ 0.5\hspace{0.25em}|1,0{>}
- 0.5\hspace{0.25em}|0,1{>}
- 0.5\hspace{0.25em}|1,1{>}$

\vspace{2mm}
\noindent
Теперь применим оператор $\hat{U}_f$,
соответствующий функции $f(x)=1$:

\noindent
\texttt{qcl> U(1,x,y)}

\noindent
[2/32] $- 0.5\hspace{0.25em}|0,0{>}
- 0.5\hspace{0.25em}|1,0{>}
+ 0.5\hspace{0.25em}|0,1{>}
+ 0.5\hspace{0.25em}|1,1{>}$

\vspace{2mm}
\noindent
Применим преобразование Адамара к биту \texttt{x}:

\noindent
\texttt{qcl> Mix(x)}

\noindent
[2/32] $- 0.70711\hspace{0.25em}|0,0{>}
+ 0.70711\hspace{0.25em}|0,1{>}$

\vspace{2mm}
\noindent
Мы видим, что бит \texttt{x} находится в состоянии $|0{>}$,
как и должно быть в случае постоянной функции.
Произведём измерение бита \texttt{x},
запишем результат в переменную \texttt{r},
и напечатаем её:

\noindent
\texttt{qcl> measure x,r}

\noindent
[2/32] $- 0.70711\hspace{0.25em}|0,0{>}
+ 0.70711\hspace{0.25em}|0,1{>}$

\vspace{2mm}
\noindent
\texttt{qcl> print r}

\noindent
0

\vspace{2mm}
\noindent
Приведём память в исходное состояние,
и повторим всё для другой функции $f(x)$:

\noindent
\texttt{qcl> reset}

\noindent
[2/32] $1\hspace{0.25em}|0,0{>}$

\vspace{2mm}
\noindent
\texttt{qcl> Not(y)}

\noindent
[2/32] $1\hspace{0.25em}|0,1{>}$

\vspace{2mm}
\noindent
\texttt{qcl> Mix(y)}

\noindent
[2/32] $0.70711\hspace{0.25em}|0,0{>}
- 0.70711\hspace{0.25em}|0,1{>}$

\vspace{2mm}
\noindent
\texttt{qcl> Mix(x)}

\noindent
[2/32] $0.5\hspace{0.25em}|0,0{>}
+ 0.5\hspace{0.25em}|1,0{>}
- 0.5\hspace{0.25em}|0,1{>}
- 0.5\hspace{0.25em}|1,1{>}$

\vspace{2mm}
\noindent
\texttt{qcl> U(2,x,y)}

\noindent
[2/32] $0.5\hspace{0.25em}|0,0{>}
- 0.5\hspace{0.25em}|1,0{>}
- 0.5\hspace{0.25em}|0,1{>}
+ 0.5\hspace{0.25em}|1,1{>}$

\vspace{2mm}
\noindent
\texttt{qcl> Mix(x)}

\noindent
[2/32] $0.70711\hspace{0.25em}|1,0{>}
- 0.70711\hspace{0.25em}|1,1{>}$

\vspace{2mm}
\noindent
\texttt{qcl> measure x,r}

\noindent
[2/32] $0.70711\hspace{0.25em}|1,0{>}
- 0.70711\hspace{0.25em}|1,1{>}$

\vspace{2mm}
\noindent
Теперь бит \texttt{x} равен 1,
как и должно быть для уравновешенной функции $f(x)=x$:

\noindent
\texttt{qcl> print r}

\noindent
1

\vspace{2mm}
\noindent
Можно написать процедуру, автоматизирующую весь алгоритм.
Параметр \texttt{n} определяет,
которую функцию $f(x)$ использовать.

\noindent
\begin{flushleft}
\texttt{qcl> procedure Deutsch(int n)}\\
\texttt{\ \ \ \ \ \{ reset;}\\
\texttt{\ \ \ \ \ \ \ \ Not(y); Mix(y); Mix(x); \ /* |+> * |-> */}\\
\texttt{\ \ \ \ \ \ \ \ U(n,x,y);}\\
\texttt{\ \ \ \ \ \ \ \ Mix(x);}\\
\texttt{\ \ \ \ \ \ \ \ measure x,r; print r;}\\
\texttt{\ \ \ \ \ \}}
\end{flushleft}

\vspace{2mm}
\noindent
\texttt{qcl> Deutsch(0)}

\noindent
0\\{}
[2/32] $0.70711\hspace{0.25em}|0,0{>}
- 0.70711\hspace{0.25em}|0,1{>}$

\vspace{2mm}
\noindent
\texttt{qcl> Deutsch(1)}

\noindent
0\\{}
[2/32] $- 0.70711\hspace{0.25em}|0,0{>}
+ 0.70711\hspace{0.25em}|0,1{>}$

\vspace{2mm}
\noindent
\texttt{qcl> Deutsch(2)}

\noindent
1\\{}
[2/32] $0.70711\hspace{0.25em}|1,0{>}
- 0.70711\hspace{0.25em}|1,1{>}$

\vspace{2mm}
\noindent
\texttt{qcl> Deutsch(3)}

\noindent
1\\{}
[2/32] $- 0.70711\hspace{0.25em}|1,0{>}
+ 0.70711\hspace{0.25em}|1,1{>}$

\vspace{2mm}
\noindent
\texttt{qcl> exit}

\vspace{2mm}

Мы видим, что алгоритм действительно выдаёт 0
для первых двух функций $f(x)$ (константы 0 и 1)
и 1 для остальных (уравновешенные функции $x$ и $1-x$).

\subsection{Пример, в котором классический алгоритм имеет\\
экспоненциальную сложность, а квантовый -- линейную}

Теперь мы рассмотрим пример задачи,
для решения которой на классическом компьютере
требуется экспоненциально большое время,
а на квантовом -- линейное~\cite{Deutsch}.
Сама по себе эта задача очень искусственная,
и интереса не представляет.
Есть и более интересные задачи,
для которых известны квантовые алгоритмы,
более эффективные, чем классические
(хотя таких задач и не очень много).
Но этот алгоритм легче всего понять.

Рассмотрим обобщение задачи из предыдущего параграфа.
А именно, рассмотрим функции $f(x)$,
отображающие $n$-битовые целые числа $x$ в 1 бит:
\[ f : \{ 0, 1 \}^n \rightarrow \{ 0, 1 \}\,. \]
Кто-то загадал функцию $f$.
Разрешено загадывать только функции, принадлежащие
одному из 2 классов:
\begin{itemize}
  \item Константы (0 или 1)
  \item ``Уравновешенные'' функции, принимающие значение 0 в половине точек
  $x$ и 1 в другой половине
\end{itemize}
Мы хотим угадать, к которому из этих двух классов принадлежит функция $f$.

В классическом случае, мы вычисляем $f(0)$, $f(1)$, \dots{}
В худшем случае, все время будет получаться одно и то же, скажем, 0.
После $2^{n-1}$ вычислений функции мы всё ещё не будем знать,
константа это или уравновешенная функция.

Теперь пусть у нас есть квантовая функция
\[ \hat{U}_f |x,y{>} = |x,y \oplus f(x){>}\,. \]
Применим её к $|x{>} \otimes |{-}{>}$:
\[ \hat{U}_f |x{>} \otimes |{-}{>} = (-1)^{f(x)} |x{>} \otimes |{-}{>}\,. \]
Теперь применим её к $|{+}{>} \otimes |{-}{>}$, где
\[ |{+}{>} = \frac{1}{\sqrt{N}} \sum_{x=0}^{N-1} |x{>}\,,\quad
N = 2^n\,. \]
Получится
\[ \hat{U}_f |{+}{>} \otimes |{-}{>} = \frac{1}{\sqrt{N}}
\sum_{x=0}^{N-1} (-1)^{f(x)} |x{>} \otimes |{-}{>}\,. \]

\begin{sloppypar}
Чтобы привести $|y{>}$ в состояние $|{-}{>}$,
применяем \texttt{Not(y); Mix(y);}.
Чтобы привести $|x{>}$ в состояние $|{+}{>}$,
применяем \texttt{Mix(x);},
то есть \texttt{Mix(x[0]); Mix(x[1]);} \dots{}
Если наша функция -- константа, то
\[ \hat{U}_f |{+}{>} \otimes |{-}{>} =
(-1)^{f(0)} |{+}{>} \otimes |{-}{>}\,, \]
то есть после применения $\hat{U}_f$ регистр $x$ будет в состоянии $|{+}{>}$.
Применим опять \texttt{Mix(x);}.
Поскольку $\hat{M}^2=1$, получится $|00\ldots0{>}$.
Если наша функция уравновешенная,
то в сумме половина членов будет со знаком $+$,
а половина -- со знаком $-$.
Поэтому состояние регистра $x$ будет ортогонально к $|{+}{>}$.
После применения \texttt{Mix(x);} получится состояние,
ортогональное к $|00\ldots0{>}$,
то есть суперпозиция любых состояний типа $|10\ldots1{>}$,
кроме состояния $|00\ldots0{>}$.
Все вычисления $f(x)$ происходят одновременно,
когда начальное состояние -- суперпозиция разных $|x{>}$.
\end{sloppypar}

\vspace{2mm}
\noindent
\texttt{QCL Quantum Computation Language (32 qubits, seed 1081610812)}

\noindent
[0/32] $1\hspace{0.25em}|0{>}$

\vspace{2mm}
\noindent
Пусть регистр \texttt{x} состоит из 3 бит.

\noindent
\texttt{qcl> qureg x[3]; qureg y[1]; int r;}

\vspace{2mm}
\noindent
Оператор $\hat{U}_f$ для 3 функций $f(x)$:
\begin{itemize}
\item $n=0$ -- $f(x)=0$ ($\hat{U}_f=1$);
\item $n=1$ -- $f(x)=x_0$
\item $n=2$ -- $f(x)=x_0 \oplus x_1 \oplus x_2$
\end{itemize}
Первая из них -- константа,
остальные -- уравновешенные функции.

\noindent
\begin{flushleft}
\texttt{qcl> procedure U(int n, qureg x, qureg y)}\\
\texttt{\ \ \ \ \ \{ if n==1 \{ x[0]->y; \} \ /* f(x)=x[0] */}\\
\texttt{\ \ \ \ \ \ \ else \{ if n==2 \{ x[0]->y; x[1]->y; x[2]->y; \}\}\}}\\
\texttt{\ \ \ \ \ \ \ \ \ \ \ \ \ \ \ \ \ \ \ \ \ \ \ \ \ \ \ \ \ /* f(x)=x[0] xor x[1] xor x[2] */}
\end{flushleft}

\vspace{2mm}
\noindent
Приготовим состояние $|{+}{>}\otimes|{-}{>}$:

\noindent
\texttt{qcl> Not(y)}

\noindent
[4/32] $1\hspace{0.25em}|000,1{>}$

\vspace{2mm}
\noindent
\texttt{qcl> Mix(y)}

\noindent
[4/32] $0.70711\hspace{0.25em}|000,0{>}
- 0.70711\hspace{0.25em}|000,1{>}$

\vspace{2mm}
\noindent
\texttt{qcl> Mix(x)}

\begin{sloppypar}
\noindent
[4/32] $0.25\hspace{0.25em}|000,0{>}
+ 0.25\hspace{0.25em}|001,0{>}
+ 0.25\hspace{0.25em}|010,0{>}
+ 0.25\hspace{0.25em}|011,0{>}
+ 0.25\hspace{0.25em}|100,0{>}
+ 0.25\hspace{0.25em}|101,0{>}
+ 0.25\hspace{0.25em}|110,0{>}
+ 0.25\hspace{0.25em}|111,0{>}
- 0.25\hspace{0.25em}|000,1{>}
- 0.25\hspace{0.25em}|001,1{>}
- 0.25\hspace{0.25em}|010,1{>}
- 0.25\hspace{0.25em}|011,1{>}
- 0.25\hspace{0.25em}|100,1{>}
- 0.25\hspace{0.25em}|101,1{>}
- 0.25\hspace{0.25em}|110,1{>}
- 0.25\hspace{0.25em}|111,1{>}$
\end{sloppypar}

\vspace{2mm}
\noindent
Применим $\hat{U}_f$ для $f(x)=0$:

\noindent
\texttt{qcl> U(0,x,y); dump}

\begin{sloppypar}
\noindent
STATE: 4 / 32 qubits allocated, 28 / 32 qubits free\\
$0.25\hspace{0.25em}|0000{>}
+ 0.25\hspace{0.25em}|0001{>}
+ 0.25\hspace{0.25em}|0010{>}
+ 0.25\hspace{0.25em}|0011{>}
+ 0.25\hspace{0.25em}|0100{>}
+ 0.25\hspace{0.25em}|0101{>}
+ 0.25\hspace{0.25em}|0110{>}
+ 0.25\hspace{0.25em}|0111{>}
- 0.25\hspace{0.25em}|1000{>}
- 0.25\hspace{0.25em}|1001{>}
- 0.25\hspace{0.25em}|1010{>}
- 0.25\hspace{0.25em}|1011{>}
- 0.25\hspace{0.25em}|1100{>}
- 0.25\hspace{0.25em}|1101{>}
- 0.25\hspace{0.25em}|1110{>}
- 0.25\hspace{0.25em}|1111{>}$
\end{sloppypar}

\vspace{2mm}
\noindent
Применим преобразование Адамара к регистру \texttt{x}:

\noindent
\texttt{qcl> Mix(x)}

\noindent
[4/32] $0.70711\hspace{0.25em}|000,0{>}
- 0.70711\hspace{0.25em}|000,1{>}$

\vspace{2mm}
\noindent
Произведём измерение регистра \texttt{x} и напечатаем результат:

\noindent
\texttt{qcl> measure x,r}

\noindent
[4/32] $0.70711\hspace{0.25em}|000,0{>}
- 0.70711\hspace{0.25em}|000,1{>}$

\vspace{2mm}
\noindent
\texttt{qcl> print r}

\noindent
0

\vspace{2mm}
\noindent
Получился 0, как и должно быть для функции $f(x)$,
являющейся константой.
Теперь очистим память и повторим всё для $f(x)=x_0$:

\noindent
\texttt{qcl> reset}

\noindent
[4/32] $1\hspace{0.25em}|000,0{>}$

\vspace{2mm}
\noindent
\texttt{qcl> Not(y); Mix(y); Mix(x)}

\begin{sloppypar}
\noindent
[4/32] $0.25\hspace{0.25em}|000,0{>}
+ 0.25\hspace{0.25em}|001,0{>}
+ 0.25\hspace{0.25em}|010,0{>}
+ 0.25\hspace{0.25em}|011,0{>}
+ 0.25\hspace{0.25em}|100,0{>}
+ 0.25\hspace{0.25em}|101,0{>}
+ 0.25\hspace{0.25em}|110,0{>}
+ 0.25\hspace{0.25em}|111,0{>}
- 0.25\hspace{0.25em}|000,1{>}
- 0.25\hspace{0.25em}|001,1{>}
- 0.25\hspace{0.25em}|010,1{>}
- 0.25\hspace{0.25em}|011,1{>}
- 0.25\hspace{0.25em}|100,1{>}
- 0.25\hspace{0.25em}|101,1{>}
- 0.25\hspace{0.25em}|110,1{>}
- 0.25\hspace{0.25em}|111,1{>}$
\end{sloppypar}

\vspace{2mm}
\noindent
\texttt{qcl> U(1,x,y)}

\begin{sloppypar}
\noindent
[4/32] $0.25\hspace{0.25em}|000,0{>}
- 0.25\hspace{0.25em}|001,0{>}
+ 0.25\hspace{0.25em}|010,0{>}
- 0.25\hspace{0.25em}|011,0{>}
+ 0.25\hspace{0.25em}|100,0{>}
- 0.25\hspace{0.25em}|101,0{>}
+ 0.25\hspace{0.25em}|110,0{>}
- 0.25\hspace{0.25em}|111,0{>}
- 0.25\hspace{0.25em}|000,1{>}
+ 0.25\hspace{0.25em}|001,1{>}
- 0.25\hspace{0.25em}|010,1{>}
+ 0.25\hspace{0.25em}|011,1{>}
- 0.25\hspace{0.25em}|100,1{>}
+ 0.25\hspace{0.25em}|101,1{>}
- 0.25\hspace{0.25em}|110,1{>}
+ 0.25\hspace{0.25em}|111,1{>}$
\end{sloppypar}

\vspace{2mm}
\noindent
\texttt{qcl> Mix(x)}

\noindent
[4/32] $0.70711\hspace{0.25em}|001,0{>}
- 0.70711\hspace{0.25em}|001,1{>}$

\vspace{2mm}
\noindent
\texttt{qcl> measure x,r; print r}

\noindent
1\\{}
[4/32] $0.70711\hspace{0.25em}|001,0{>}
- 0.70711\hspace{0.25em}|001,1{>}$

\vspace{2mm}
\noindent
Результат ненулевой,
как и должно быть для уравновешенной функции.
Повторим ещё раз для $f(x)=x_0 \oplus x_1 \oplus x_2$:

\noindent
\texttt{qcl> reset}

\noindent
[4/32] $1\hspace{0.25em}|000,0{>}$

\vspace{2mm}
\noindent
\texttt{qcl> Not(y); Mix(y); Mix(x)}

\begin{sloppypar}
\noindent
[4/32] $0.25\hspace{0.25em}|000,0{>}
+ 0.25\hspace{0.25em}|001,0{>}
+ 0.25\hspace{0.25em}|010,0{>}
+ 0.25\hspace{0.25em}|011,0{>}
+ 0.25\hspace{0.25em}|100,0{>}
+ 0.25\hspace{0.25em}|101,0{>}
+ 0.25\hspace{0.25em}|110,0{>}
+ 0.25\hspace{0.25em}|111,0{>}
- 0.25\hspace{0.25em}|000,1{>}
- 0.25\hspace{0.25em}|001,1{>}
- 0.25\hspace{0.25em}|010,1{>}
- 0.25\hspace{0.25em}|011,1{>}
- 0.25\hspace{0.25em}|100,1{>}
- 0.25\hspace{0.25em}|101,1{>}
- 0.25\hspace{0.25em}|110,1{>}
- 0.25\hspace{0.25em}|111,1{>}$
\end{sloppypar}

\vspace{2mm}
\noindent
\texttt{qcl> U(2,x,y)}

\begin{sloppypar}
\noindent
[4/32] $0.25\hspace{0.25em}|000,0{>}
- 0.25\hspace{0.25em}|001,0{>}
- 0.25\hspace{0.25em}|010,0{>}
+ 0.25\hspace{0.25em}|011,0{>}
- 0.25\hspace{0.25em}|100,0{>}
+ 0.25\hspace{0.25em}|101,0{>}
+ 0.25\hspace{0.25em}|110,0{>}
- 0.25\hspace{0.25em}|111,0{>}
- 0.25\hspace{0.25em}|000,1{>}
+ 0.25\hspace{0.25em}|001,1{>}
+ 0.25\hspace{0.25em}|010,1{>}
- 0.25\hspace{0.25em}|011,1{>}
+ 0.25\hspace{0.25em}|100,1{>}
- 0.25\hspace{0.25em}|101,1{>}
- 0.25\hspace{0.25em}|110,1{>}
+ 0.25\hspace{0.25em}|111,1{>}$
\end{sloppypar}

\vspace{2mm}
\noindent
\texttt{qcl> Mix(x)}

\noindent
[4/32] $0.70711\hspace{0.25em}|111,0{>}
- 0.70711\hspace{0.25em}|111,1{>}$

\vspace{2mm}
\noindent
\texttt{qcl> measure x,r; print r}

\noindent
7\\{}
[4/32] $0.70711\hspace{0.25em}|111,0{>}
- 0.70711\hspace{0.25em}|111,1{>}$

\vspace{2mm}
\noindent
Результат ненулевой, как и должно быть.

\noindent
\texttt{qcl> exit}

Какова сложность этого алгоритма?
\texttt{Mix(x);} означает \texttt{Mix(x[0]); Mix(x[1]);} \dots{}
Мы делаем \texttt{Mix(x);} дважды, то есть $2n$ элементарных операций.
Сложность линейная вместо экспоненциальной!

Все известные классические алгоритмы факторизации
больших целых чисел ($n$-битовых) имеют сложность,
растущую быстрее любой степени $n$.
В 1994 году Питер Шор~\cite{Shor} предложил квантовый алгоритм,
имеющий полиномиальную сложность.
Многие популярные программы (ssh, pgp и т.д.)
используют криптосистему с публичными ключами RSA~\cite{RSA}.
Публичный ключ содержит большое целое число;
если факторизовать его, можно легко найти приватный ключ.
Безопасность интернета держится на том,
что известные алгоритмы не позволяют факторизовать такие числа
за обозримое время.
Если появится устройство, позволяющее факторизовать такие числа
достаточно быстро, то тот, у кого оно есть,
сможет читать зашифрованные письма, подсматривать пароли и т.д.

Эмулятор QCL делает вычисления
с разными компонентами состояний последовательно,
а не одновременно, как настоящий квантовый компьютер.
Поэтому его скорость экспоненциально убывает с ростом числа битов.
Естественно, эмулятор не годится для того, чтобы получать выгоду
от быстрых квантовых алгоритмов;
для этого нужен настоящий квантовый компьютер.

\section{Квантовая криптография}

Раньше в литературе по криптографии говорили что-нибудь вроде:
``A посылает сообщение B''.
В последнее время вместо этого принято говорить:
``Алиса посылает сообщение Бобу''.
В трёхсторонних обменах мнениями участвует также Сесиль
(девичья фамилия C).
Есть также зловредный персонаж Ева,
которая занимается подслушиванием (evesdropping).

Алиса и Боб хотят передавать друг другу сообщения
таким образом, что, если их перехватит Ева,
то она не сможет их расшифровать.
Для этого существует надёжный метод --
использование одноразовых блокнотов.
Допустим, что у Алисы и Боба есть одинаковые блокноты,
на страничках которых написана случайная последовательность 0 и 1.
Больше ни у кого во всём мире этой последовательности нет.
Сообщения -- это файлы, т.е.\ последовательности битов.
Алиса прибавляет по модулю 2 (т.е.\ делает Xor)
к битам своего файла биты из блокнота.
Использованные странички она вырывает и сжигает.
Зашифрованный файл она посылает Бобу по email.
Боб прибавляет по модулю 2 (т.е.\ делает Xor)
к полученным битам биты из своего блокнота.
Использованные странички он вырывает и сжигает.
После этого он читает расшифрованный файл.
Если email перехватила Ева,
она никак не сможет расшифровать сообщение
(предполагается, что одна и та же последовательность
никогда не используется дважды).

Если Алиса работает резидентом в тылу врага,
то передать ей из рук в руки новый блокнот взамен закончившегося
может быть затруднительно.
Посылать эту случайную последовательность битов
по классическому кабелю любой природы опасно --
Ева может скопировать их, и сможет расшифровывать сообщения.

Простой метод, основанный на квантовой механике,
предложен в~\cite{Crypto}.
Пусть кто-то создаёт EPR-состояния пар частиц со спином $\frac{1}{2}$
с суммарным спином 0.
Одну частицу посылают Алисе, а другую Бобу.
Алиса кидает монетку и решает, измерять ей $z$-компоненту спина
или $x$-компоненту; Боб делает то же самое.
Они вывешивают на своих home-page на WWW (где все могут видеть)
последовательность орлов и решек,
и выбрасывают бесполезные результаты измерений,
в которых они мерили разные компоненты (примерно 50\%).
Остальные результаты дают случайную последовательность 0 и 1,
одинаковую у Алисы и Боба.

Теперь допустим, что Ева перехватывает частицы, посылаемые Алисе.
Она не знает, какую компоненту спина ей нужно мерить,
и в 50\%  случаев меряет не ту.
В этих случаях Алиса и Боб получают не EPR-пару,
а нескоррелированную пару спинов.
То есть примерно 25\% результатов Алисы и Боба отличаются.
Алиса может послать Бобу открытым текстом некоторую часть своих 0 и 1
(эту часть они использовать не будут).
Если Боб обнаружит 25\%  расхождений со своими,
то они будут знать, что их подслушивали.
Если же нет, то в это время их не подслушивали,
и есть надежда, что и в остальное время тоже.

\vspace{2mm}
\noindent
\texttt{QCL Quantum Computation Language (32 qubits, seed 1081653831)}

\noindent
[0/32] $1\hspace{0.25em}|0{>}$

\vspace{2mm}
\noindent
Пара квантовых бит:

\noindent
\texttt{qcl> qureg x[2]}

\vspace{2mm}
\noindent
Процедура, создающая EPR-состояние с суммарным спином 0:

\noindent
\begin{flushleft}
\texttt{qcl> operator EPR(qureg x)}\\
\texttt{\ \ \ \ \ \{ Not(x[0]); Mix(x[0]); x[0]->x[1]; Not(x[0]); \}}
\end{flushleft}

\vspace{2mm}
\noindent
Проверим:

\noindent
\texttt{qcl> EPR(x)}

\noindent
[2/32] $0.70711\hspace{0.25em}|01{>}
- 0.70711\hspace{0.25em}|10{>}$

\vspace{2mm}
\noindent
Эта процедура описывает генерацию и измерение $n$ EPR-пар.
Если у Алисы выпал орёл, она записывает это (\texttt{mA=1})
и поворачивает свой спин на $\pi/2$ вокруг оси $x$.
Результат измерения она записывает в \texttt{rA}.
Аналогично, у Боба тип измерения записывается в \texttt{mB},
а результат в \texttt{rB}.
Если \texttt{mA} и \texttt{mB} совпадают, результаты принимаются.

\noindent
\begin{flushleft}
\texttt{qcl> procedure Crypto(int n, boolean Eve)}\\
\texttt{\ \ \ \ \ \{ int mA; int rA; int mB; int rB; int i;}\\
\texttt{\ \ \ \ \ \ \ for i=1 to n}\\
\texttt{\ \ \ \ \ \ \ \{ reset; EPR(x);}\\
\texttt{\ \ \ \ \ \ \ \ \ /* Eve */}\\
\texttt{\ \ \ \ \ \ \ \ \ if Eve \{ measure x[0]; \}}\\
\texttt{\ \ \ \ \ \ \ \ \ /* Alice */}\\
\texttt{\ \ \ \ \ \ \ \ \ if random()>0.5 \{ mA=1; RotX(pi/2,x[0]); \}}\\
\texttt{\ \ \ \ \ \ \ \ \ else \{ mA=0; \}}\\
\texttt{\ \ \ \ \ \ \ \ \ measure x[0],rA;}\\
\texttt{\ \ \ \ \ \ \ \ \ /* Bob */}\\
\texttt{\ \ \ \ \ \ \ \ \ if random()>0.5 \{ mB=1; RotX(pi/2,x[1]); \}}\\
\texttt{\ \ \ \ \ \ \ \ \ else \{ mB=0; \}}\\
\texttt{\ \ \ \ \ \ \ \ \ measure x[1],rB;}\\
\texttt{\ \ \ \ \ \ \ \ \ rB=1-rB;}\\
\texttt{\ \ \ \ \ \ \ \ \ /* Comparison */}\\
\texttt{\ \ \ \ \ \ \ \ \ if mA==mB \{ print rA,rB; \}}\\
\texttt{\ \ \ \ \ \ \ \}}\\
\texttt{\ \ \ \ \ \}}
\end{flushleft}

\vspace{2mm}
\noindent
Если нет подслушивания, результаты должны совпадать:

\noindent
\texttt{qcl> Crypto(100,false)}

\noindent
0 0\hspace{1cm}
1 1\hspace{1cm}
1 1\hspace{1cm}
1 1\hspace{1cm}
0 0\hspace{1cm}
1 1\hspace{1cm}
1 1\hspace{1cm}
1 1\\
0 0\hspace{1cm}
1 1\hspace{1cm}
0 0\hspace{1cm}
1 1\hspace{1cm}
1 1\hspace{1cm}
0 0\hspace{1cm}
0 0\hspace{1cm}
0 0\\
1 1\hspace{1cm}
1 1\hspace{1cm}
0 0\hspace{1cm}
1 1\hspace{1cm}
1 1\hspace{1cm}
0 0\hspace{1cm}
1 1\hspace{1cm}
0 0\\
0 0\hspace{1cm}
1 1\hspace{1cm}
0 0\hspace{1cm}
0 0\hspace{1cm}
1 1\hspace{1cm}
1 1\hspace{1cm}
1 1\hspace{1cm}
0 0\\
1 1\hspace{1cm}
1 1\hspace{1cm}
1 1\hspace{1cm}
0 0\hspace{1cm}
0 0\hspace{1cm}
1 1\hspace{1cm}
1 1\hspace{1cm}
1 1\\
1 1\hspace{1cm}
0 0\hspace{1cm}
0 0\hspace{1cm}
0 0\hspace{1cm}
1 1\hspace{1cm}
0 0\hspace{1cm}
1 1\\{}
[2/32] $1\hspace{0.25em}|01{>}$

\vspace{2mm}
\noindent
Если параметр \texttt{Eve} есть \texttt{true},
то Ева измеряет $z$-компоненту спина частиц, посылаемых Алисе.
В этом случае примерно в 25\% результатов обнаружатся расхождения:

\noindent
\texttt{qcl> Crypto(100,true)}

\noindent
0 0\hspace{1cm}
0 0\hspace{1cm}
1 0\hspace{1cm}
0 0\hspace{1cm}
0 0\hspace{1cm}
1 0\hspace{1cm}
0 1\hspace{1cm}
1 1\\
0 0\hspace{1cm}
1 1\hspace{1cm}
1 1\hspace{1cm}
1 0\hspace{1cm}
0 0\hspace{1cm}
1 1\hspace{1cm}
0 1\hspace{1cm}
1 1\\
0 1\hspace{1cm}
0 1\hspace{1cm}
1 0\hspace{1cm}
0 0\hspace{1cm}
1 1\hspace{1cm}
0 1\hspace{1cm}
1 1\hspace{1cm}
0 0\\
1 1\hspace{1cm}
0 1\hspace{1cm}
0 0\hspace{1cm}
0 0\hspace{1cm}
0 0\hspace{1cm}
0 0\hspace{1cm}
0 1\hspace{1cm}
1 1\\
1 1\hspace{1cm}
1 1\hspace{1cm}
1 1\hspace{1cm}
1 1\hspace{1cm}
1 1\hspace{1cm}
0 0\hspace{1cm}
0 0\hspace{1cm}
0 0\\
0 0\hspace{1cm}
1 1\\{}
[2/32] $- i\hspace{0.25em}|11{>}$

\vspace{2mm}
\noindent
\texttt{qcl> exit}

\vspace{2mm}
Похожая идея -- деньги, защищённые от подделки
принципами квантовой механики~\cite{Money}.
Выпускающий их банк наносит на каждую банкноту номер,
а также встраивает несколько десятков спинов $\frac{1}{2}$.
Каждый из них имеет определённую проекцию на $z$ или $x$.
В базе данных банка для каждого номера банкноты хранится информация о том,
какие спины имеют определённую $\hat{s}_z$, какие -- $\hat{s}_x$,
и чему именно они равны.
(В исходной формулировке речь шла про фотоны,
запертые между параллельными зеркалами
и имеющие определённую линейную либо циркулярную поляризацию;
это, конечно, несущественно.)
Служащий банка может проверить подлинность банкноты,
измерив нужные проекции спинов.
Фальшивомонетчик не знает, какую проекцию измерять для какого спина,
и в половине случаев будет измерять не ту.
Поэтому он не только не сделает хорошей копии
(которая выдержала бы проверку),
но и испортит оригинал.

\section{Квантовая телепортация}

Пусть у Алисы есть частица со спином $\frac{1}{2}$,
находящаяся в некотором состоянии
\[ |x{>} = \alpha |0{>} + \beta |1{>}\,. \]
Кто-то готовит пару частиц со спином $\frac{1}{2}$
в EPR-состоянии
\[ |y{>} = \frac{|00{>} + |11{>}}{\sqrt{2}}\,, \]
и посылает первую из них Алисе, а вторую Бобу.
Тогда можно привести частицу Боба в точно то же состояние
$\alpha |0{>} + \beta |1{>}$
(при этом бит Алисы уже не будет в этом состоянии)~\cite{Tele}.

Итак, начальное состояние
\begin{eqnarray*}
&& \left(\alpha|0{>} + \beta|1{>}\right) \otimes
\frac{|00{>} + |11{>}}{\sqrt{2}}\\
&&{} = \frac{1}{\sqrt{2}}
\left[\alpha \left(|000{>}+|011{>}\right)
+ \beta \left(|100{>} + |111{>}\right) \right]\,.
\end{eqnarray*}
Алиса применяет к своей паре бит $\hat{C}$:
\[ \frac{1}{\sqrt{2}} \left[\alpha \left(|000{>} + |011{>}\right)
+ \beta \left(|110{>} + |101{>}\right) \right]\,, \]
затем к первому биту $\hat{M}$:
\begin{eqnarray*}
&& \frac{\alpha}{2} \left(|000{>} + |100{>} + |011{>} + |111{>}\right)\\
&&{} + \frac{\beta}{2} \left(|010{>} - |110{>} + |001{>} - |101{>}\right)\\
&&{} = \frac{1}{2} \left(|00{>} \otimes |x{>} + |01{>} \otimes \sigma_x |x{>}
+ |10{>} \otimes \sigma_z |x{>} - i|11{>} \otimes \sigma_y |x{>} \right)\,,
\end{eqnarray*}
где
\[ \sigma_x = \left(\begin{array}{cc}
     0 & 1\\
     1 & 0
   \end{array}\right)\,,\quad \sigma_y = \left(\begin{array}{cc}
     0 & - i\\
     i & 0
   \end{array}\right)\,,\quad \sigma_z = \left(\begin{array}{cc}
     1 & 0\\
     0 & - 1
   \end{array}\right) \]
-- матрицы Паули (они унитарны).
Алиса измеряет 2 своих бита, и получает классическое двухбитовое число
(0, 1, 2 или 3 с равными вероятностями).
Она посылает это число Бобу по классическому каналу (скажем, по email).
Боб действует на свой бит оператором $1$, $\sigma_x$, $\sigma_z$
или $-i\sigma_y$, соответственно.

\vspace{2mm}
\noindent
\texttt{QCL Quantum Computation Language (32 qubits, seed 1087133710)}

\noindent
[0/32] $1\hspace{0.25em}|0{>}$

\vspace{2mm}
\noindent
\texttt{x} -- бит Алисы;
\texttt{y} -- пара бит для EPR-состояния

\noindent
\texttt{qcl> qureg x[1]; qureg y[2]; int r}

\vspace{2mm}
\noindent
Процедура, создающая EPR-состояние:

\noindent
\begin{flushleft}
\texttt{qcl> operator EPR(qureg y)}\\
\texttt{\ \ \ \ \ \{ Mix(y[0]); y[0]->y[1]; \}}
\end{flushleft}

\vspace{2mm}
\noindent
Проверим:

\noindent
\texttt{qcl> EPR(y)}

\noindent
[3/32] $0.70711\hspace{0.25em}|0,00{>}
+ 0.70711\hspace{0.25em}|0,11{>}$

\vspace{2mm}
\noindent
\texttt{qcl> reset}

\noindent
[3/32] $1\hspace{0.25em}|0,00{>}$

\vspace{2mm}
\noindent
Приведём бит Алисы в случайное состояние:

\noindent
\texttt{qcl> randomize(x)}

\noindent
[3/32] $(-0.35679-0.77939i)\hspace{0.25em}|0,00{>}
+ (0.50955+0.074842i)\hspace{0.25em}|1,00{>}$

\vspace{2mm}
\noindent
Приготовим EPR-состояние пары \texttt{y}:

\noindent
\texttt{qcl> EPR(y)}

\begin{sloppypar}
\noindent
[3/32] $(-0.25229-0.55111i)\hspace{0.25em}|0,00{>}
+ (0.36031+0.052921i)\hspace{0.25em}|1,00{>}
+ (-0.25229-0.55111i)\hspace{0.25em}|0,11{>}
+ (0.36031+0.052921i)\hspace{0.25em}|1,11{>}$
\end{sloppypar}

\vspace{2mm}
\noindent
Алиса применяет \texttt{CNot} к своим 2 битам:

\noindent
\texttt{qcl> x->y[1]}

\begin{sloppypar}
\noindent
[3/32] $(-0.25229-0.55111i)\hspace{0.25em}|0,00{>}
+ (0.36031+0.052921i)\hspace{0.25em}|1,01{>}
+ (0.36031+0.052921i)\hspace{0.25em}|1,10{>}
+ (-0.25229-0.55111i)\hspace{0.25em}|0,11{>}$
\end{sloppypar}

\vspace{2mm}
\noindent
Алиса применяет преобразование Адамара к биту \texttt{x}:

\noindent
\texttt{qcl> Mix(x)}

\begin{sloppypar}
\noindent
[3/32] $(-0.1784-0.3897i)\hspace{0.25em}|0,00{>}
+ (-0.1784-0.3897i)\hspace{0.25em}|1,00{>}
+ (0.25478+0.037421i)\hspace{0.25em}|0,01{>}
+ (-0.25478-0.037421i)\hspace{0.25em}|1,01{>}
+ (0.25478+0.037421i)\hspace{0.25em}|0,10{>}
+ (-0.25478-0.037421i)\hspace{0.25em}|1,10{>}
+ (-0.1784-0.3897i)\hspace{0.25em}|0,11{>}
+ (-0.1784-0.3897i)\hspace{0.25em}|1,11{>}$
\end{sloppypar}

\vspace{2mm}
\noindent
Алиса измеряет свои 2 бита:

\noindent
\texttt{qcl> measure y[1]\&x,r; print r}

\noindent
1\\{}
[3/32] $(0.50955+0.074842i)\hspace{0.25em}|0,10{>}
+ (-0.35679-0.77939i)\hspace{0.25em}|0,11{>}$

\vspace{2mm}
\noindent
Боб действует на свой бит \texttt{y[0]}
в соответствии с полученной информацией:

\noindent
\begin{flushleft}
\texttt{qcl> if r==1 \{ X(y[0]); \}}\\
\texttt{\ \ \ \ \ else \{ if r==2 \{ Z(y[0]); \}}\\
\texttt{\ \ \ \ \ else \{ if r==3 \{ RotY(-pi,y[0]); \}\}\}}
\end{flushleft}

\noindent
[3/32] $(-0.35679-0.77939i)\hspace{0.25em}|0,10{>}
+ (0.50955+0.074842i)\hspace{0.25em}|0,11{>}$

\vspace{2mm}
\noindent
Теперь бит \texttt{y[0]}, находящийся у Боба,
находится в состоянии $|x{>}$.

\vspace{2mm}
\noindent
Процедура \texttt{Tele} повторяет это \texttt{n} раз.
Она печатает состояние перед телепортацией и после неё.

\noindent
\begin{flushleft}
\texttt{qcl> procedure Tele(int n)}\\
\texttt{\ \ \ \ \ \{ int i;}\\
\texttt{\ \ \ \ \ \ \ for i=1 to n}\\
\texttt{\ \ \ \ \ \ \ \{ reset;}\\
\texttt{\ \ \ \ \ \ \ \ \ randomize(x); dump;}\\
\texttt{\ \ \ \ \ \ \ \ \  EPR(y);}\\
\texttt{\ \ \ \ \ \ \ \ \ /* Alice */}\\
\texttt{\ \ \ \ \ \ \ \ \ x->y[1]; Mix(x);}\\
\texttt{\ \ \ \ \ \ \ \ \ measure y[1]\&x,r; print r;}\\
\texttt{\ \ \ \ \ \ \ \ \ /* Bob */}\\
\texttt{\ \ \ \ \ \ \ \ \ if r==1 \{ X(y[0]); \}}\\
\texttt{\ \ \ \ \ \ \ \ \ else \{ if r==2 \{ Z(y[0]); \}}\\
\texttt{\ \ \ \ \ \ \ \ \ else \{ if r==3 \{ RotY(-pi,y[0]); \}\}\}}\\
\texttt{\ \ \ \ \ \ \ \ \ dump;}\\
\texttt{\ \ \ \ \ \ \ \}}\\
\texttt{\ \ \ \ \ \}}
\end{flushleft}

\vspace{2mm}
\noindent
\texttt{qcl> Tele(10)}

\begin{sloppypar}
\vspace{2mm}
\noindent
STATE: 3 / 32 qubits allocated, 29 / 32 qubits free\\
$(0.25304+0.34095i)\hspace{0.25em}|000{>}
+ (-0.41323+0.80558i)\hspace{0.25em}|001{>}$\\
3\\
STATE: 3 / 32 qubits allocated, 29 / 32 qubits free\\
$(0.25304+0.34095i)\hspace{0.25em}|101{>}
+ (-0.41323+0.80558i)\hspace{0.25em}|111{>}$\\[2mm]
STATE: 3 / 32 qubits allocated, 29 / 32 qubits free\\
$(0.25496-0.77909i)\hspace{0.25em}|000{>}
+ (0.38057+0.42799i)\hspace{0.25em}|001{>}$\\
0\\
STATE: 3 / 32 qubits allocated, 29 / 32 qubits free\\
$(0.25496-0.77909i)\hspace{0.25em}|000{>}
+ (0.38057+0.42799i)\hspace{0.25em}|010{>}$\\[2mm]
STATE: 3 / 32 qubits allocated, 29 / 32 qubits free\\
$(0.95379+0.28287i)\hspace{0.25em}|000{>}
+ (0.06279-0.079523i)\hspace{0.25em}|001{>}$\\
2\\
STATE: 3 / 32 qubits allocated, 29 / 32 qubits free\\
$(0.95379+0.28287i)\hspace{0.25em}|001{>}
+ (0.06279-0.079523i)\hspace{0.25em}|011{>}$\\[2mm]
STATE: 3 / 32 qubits allocated, 29 / 32 qubits free\\
$(-0.74984+0.16696i)\hspace{0.25em}|000{>}
+ (0.27132+0.57987i)\hspace{0.25em}|001{>}$\\
2\\
STATE: 3 / 32 qubits allocated, 29 / 32 qubits free\\
$(-0.74984+0.16696i)\hspace{0.25em}|001{>}
+ (0.27132+0.57987i)\hspace{0.25em}|011{>}$\\[2mm]
STATE: 3 / 32 qubits allocated, 29 / 32 qubits free\\
$(0.19238-0.22205i)\hspace{0.25em}|000{>}
+ (0.61016+0.73579i)\hspace{0.25em}|001{>}$\\
2\\
STATE: 3 / 32 qubits allocated, 29 / 32 qubits free\\
$(0.19238-0.22205i)\hspace{0.25em}|001{>}
+ (0.61016+0.73579i)\hspace{0.25em}|011{>}$\\[2mm]
STATE: 3 / 32 qubits allocated, 29 / 32 qubits free\\
$(-0.074875-0.17213i)\hspace{0.25em}|000{>}
+ (-0.98212-0.014345i)\hspace{0.25em}|001{>}$\\
3\\
STATE: 3 / 32 qubits allocated, 29 / 32 qubits free\\
$(-0.074875-0.17213i)\hspace{0.25em}|101{>}
+ (-0.98212-0.014345i)\hspace{0.25em}|111{>}$\\[2mm]
STATE: 3 / 32 qubits allocated, 29 / 32 qubits free\\
$(-0.13448-0.52058i)\hspace{0.25em}|000{>}
+ (-0.68431+0.49257i)\hspace{0.25em}|001{>}$\\
2\\
STATE: 3 / 32 qubits allocated, 29 / 32 qubits free\\
$(-0.13448-0.52058i)\hspace{0.25em}|001{>}
+ (-0.68431+0.49257i)\hspace{0.25em}|011{>}$\\[2mm]
STATE: 3 / 32 qubits allocated, 29 / 32 qubits free\\
$(-0.042565-0.40042i)\hspace{0.25em}|000{>}
+ (-0.27303-0.87367i)\hspace{0.25em}|001{>}$\\
3\\
STATE: 3 / 32 qubits allocated, 29 / 32 qubits free\\
$(-0.042565-0.40042i)\hspace{0.25em}|101{>}
+ (-0.27303-0.87367i)\hspace{0.25em}|111{>}$\\[2mm]
STATE: 3 / 32 qubits allocated, 29 / 32 qubits free\\
$(0.0019455+0.0098884i)\hspace{0.25em}|000{>}
+ (0.90142+0.43282i)\hspace{0.25em}|001{>}$\\
1\\
STATE: 3 / 32 qubits allocated, 29 / 32 qubits free\\
$(0.0019455+0.0098884i)\hspace{0.25em}|100{>}
+ (0.90142+0.43282i)\hspace{0.25em}|110{>}$\\[2mm]
STATE: 3 / 32 qubits allocated, 29 / 32 qubits free\\
$(0.77744-0.49842i)\hspace{0.25em}|000{>}
+ (-0.11673-0.36542i)\hspace{0.25em}|001{>}$\\
2\\
STATE: 3 / 32 qubits allocated, 29 / 32 qubits free\\
$(0.77744-0.49842i)\hspace{0.25em}|001{>}
+ (-0.11673-0.36542i)\hspace{0.25em}|011{>}$\\[2mm]
[3/32] $(0.77744-0.49842i)\hspace{0.25em}|1,00{>}
+ (-0.11673-0.36542i)\hspace{0.25em}|1,01{>}$
\end{sloppypar}

\noindent
\texttt{qcl> exit}

\vspace{2mm}
На первый взгляд это кажется странным.
Состояние спина у Алисы описывается двумя комплексными числами
$\alpha$ и $\beta$, в которых, вообще говоря,
бесконечно много значащих цифр (десятичных или двоичных).
Алиса посылает Бобу всего 2 бита информации,
и это состояние передаётся спину у Боба.
На самам деле Алиса не знала состояния спина
(т.е.\ $\alpha$ и $\beta$),
так что никакой передачи информации здесь нет
(если бы она знала, она могла бы просто послать Бобу
email с инструкцией, как это состояние приготовить).

Если представить себе возможность телепортации квантового состояния
более сложных систем, чем спин $\frac{1}{2}$, например, человека,
то это бы выглядело так.
В пункте A человек заходит в телепортационную кабину.
Результаты измерений передаются в пункт B по email.
Там из телепортационной кабины выходит точно такой же человек.
В первой кабине вместо человека остаётся груда мусора
(потому что состояние первой системы не сохраняется).

\end{document}